# Equivalent Relation between Normalized Spatial Entropy and Fractal Dimension

Yanguang Chen

(Department of Geography, College of Urban and Environmental Sciences, Peking University, Beijing 100871, P.R. China. E-mail: chenyg@pku.edu.cn)

**Abstract**: Fractal dimension is defined on the base of entropy, including macro state entropy and information entropy. The generalized correlation dimension of multifractals is based on Renyi entropy. However, the mathematical transform from entropy to fractal dimension is not yet clear in both theory and practice. This paper is devoted to revealing the new equivalence relation between spatial entropy and fractal dimension using functional box-counting method. Based on varied regular fractals, the numerical relationship between spatial entropy and fractal dimension is examined. The results show that the ratio of actual entropy ($M_q$) to the maximum entropy ($M_{max}$) equals the ratio of actual dimension ($D_q$) to the maximum dimension ($D_{max}$). The spatial entropy and fractal dimension of complex spatial systems can be converted into one another by means of functional box-counting method. The theoretical inference is verified by observational data of urban form. A conclusion is that normalized spatial entropy is equal to normalized fractal dimension. Fractal dimensions proved to be the characteristic values of entropies. In empirical studies, if the linear size of spatial measurement is small enough, a normalized entropy value is infinitely approximate to the corresponding normalized fractal dimension value. Based on the theoretical result, new spatial measurements of urban space filling can be defined, and multifractal parameters can be generalized to describe both simple systems and complex systems.

# 1. Introduction

Entropy is one of important concepts in modern science, related with the processes and patterns of time, space, and information. Today, entropy is often associated with fractals and fractal dimension. Fractals are as important to future science what entropy is to modern science (Wheeler, 1983). Both entropy and fractal dimension are measures of complexity (Bak, 1996; Cramer, 1993; Pincus, 1991), especially, entropy can be used to measure multiscale complexity (Bar-Yam, 2004a; Bar-Yam, 2004b). A problem is that entropy values always depend on the scale of measurement. If we change the linear scale for measurement, the entropy value will change, and this results in uncertainty. In contrast, fractal dimension is independent of measurement scales. Varied fractal dimension formulae are actually defined based on entropy functions. There are inherent relationships between entropy and fractal dimension. The generalized dimension of multifractals is based on Renyi entropy. In the multifractal dimension set, capacity dimension is based on macro state entropy, and information dimension is based on Shannon's information entropy (Feder, 1988). It was also demonstrated that there are analogies between multifractals and thermodynamics, and the Legendre transform is an analogue of entropy (Stanley and Meakin, 1988). In fact, Shannon's entropy proved to be equivalent in mathematics to Hausdorff dimension and Kolmogorov complexity (Ryabko, 1986). For regular fractals without overlapped fractal copies, *Hausdorff dimension* equals *similarity dimension* and can be replaced by *box dimension*.

Complexity science has been applied to urban and regional studies. Cities and regions are multiscale complex spatial systems (Allen, 1997; Chen, 2008; Wilson, 2000), which cannot be understood by the ideas from reductionism. Entropy is one measure of spatial complexity (Wilson, 2000; Wilson, 2010), and fractal dimension is another measure of spatial complexity (Batty, 2005; Batty and Longley, 1994; Frankhauser, 1994; White and Engelen, 1994). In order to make spatial analysis of cities and regions, the theory of spatial entropy came into being (Batty, 1974; Batty, 1976; Batty, 2010; Batty *et al*, 2014). One of the obstacles to developing the spatial entropy theory, as indicated above, lies in dependence on measurement scales. Spatial entropy can be calculated based on geographical zoning network (Batty and Sammons, 1979). However, geographical zoning often causes the modifiable areal unit problem (MAUP) (Cressie, 1996; Kwan, 2012; Openshaw, 1983; Unwin, 1996). The sizes and scales of areal units influence the entropy values. If we transform spatial entropy into fractal dimension, the problem can be solved. However, entropy cannot be converted into fractal dimension by geographical zoning. A discovery is that the box-counting method can be employed to measure both spatial entropy and fractal dimension. This is an important approach for geographical spatial analysis. First, by box counting, the entropy concept can be easily generalized to the notion of spatial entropy. Second, by box counting, the mathematical relationships and distinction between entropy and fractal dimension can be illustrated clearly. Third, by box counting, both the spatial measurements of entropy and fractal dimension can be normalized, and thus it is easy to define the extreme entropy values based on a normal framework, including the maximum entropy and minimum entropy.

This paper is devoted to revealing the mathematical and numerical relationships between spatial entropy and fractal dimension. In theory, if we treat a fractal as a dynamic process rather than a static pattern, spatial entropy can be converted into fractal dimension. In practice, spatial entropy and fractal dimension of a complex system such as urban form can be associated with each other by functional box-counting methods. This method is proposed by Lovejoy *et al* (1987) and consolidated by Chen (1995). Suppose that a growing fractal system comes between two extreme states: one is absolutely concentrated state with the minimum entropy and fractal dimension (e.g., one or more points), and the other is absolutely filled state with the maximum entropy and fractal dimension (e.g., a circular or square area). Using simple regular fractals, we can demonstrate that the normalized entropy equals the normalized fractal dimension, and a fractal dimension is a characteristic value of the corresponding entropy. Thus fractal evolution and fractal dimension growth of complex systems can be understood by entropy theory. The rest parts of this article are organized as follows. In Section 2, the theoretical results are derived from simple mathematical transform. In Section 3, the result is testified using simple regular fractal analysis and empirical analysis of real city. In Section 4, the related questions are discussed and the theory is extended. Finally, the conclusions are reached on the base of the theoretical and case studies.

## 2. Results

### 2.1 The correspondence between entropy and fractal dimension

Using regular fractals and box dimension concept, we can deduce a new relationship between entropy and fractal dimension. Different types of fractal dimension are defined on the base of different entropy functions. In multifractal theory, the generalized dimension is based on Renyi entropy that can be expressed as (Renyi, 1961)

$$M_q = -\frac{1}{q-1} \ln \sum_{i=1}^{N} P_i^q,  \quad (1)$$

where $q$ denotes the order of moment ($q=\ldots,-2,-1, 0, 1, 2,\ldots$), $M_q$ represents Rengyi entropy, $P_i$ is probability ($\sum P_i=1$), and $N$ is the number of nonempty boxes ($i=1, 2,\ldots, N$). What is more, the symbol "ln" refers to natural logarithm. If the moment order $q=0$, equation (1) changes to

$$S = M_0 = \ln N, \quad (2)$$

where $S$ denotes Boltzmann's macro state entropy. The macro state entropy differs from Boltzmann's micro state entropy defined by permutation and combination of all elements. If $q=1$, equation (1) can be reduced to

$$H = M_1 = -\sum_{i=1}^{N} P_i \ln P_i, \quad (3)$$

where $H$ refers to Shannon's information entropy (Shannon, 1948). For simplicity, the proportionality coefficient is taken as 1. Where a regular simple fractal is concerned, Shannon's entropy can be regarded as micro state entropy. For the equiprobable distribution, we have $P_i=1/N$, and thus information entropy will return to the macro state entropy. If $q=2$, equation (2) can be expressed as

$$M_2 = -\ln \sum_{i=1}^{N} P_i^2, \quad (4)$$

which is the second order Renyi entropy., Because equation (4) proved to be a correlation function, $M_2$ can be treated as correlation entropy. If $P_i=1/N$ as given, then correlation entropy will return to the macro state entropy. This suggests that the three types of entropy values are equal to one another under the condition of equiprobable distribution ($P_i=1/N$).

The entropy functions can be used to describe the systems with characteristic lengths. For simple systems, an entropy value can be uniquely determined. However, for the scale-free systems such as fractal cities, the Renyi entropy is not determinate, and an alternative measurement is the generalized fractal dimension. Based on Renyi entropy and box-counting method, the generalized dimension is defined as below (Feder, 1988; Grassberger1985; Mandelbrot, 1999)

$$D_q = -\frac{M_q(\varepsilon)}{\ln \varepsilon} = -\lim_{\varepsilon \to 0} \frac{1}{q-1} \frac{\ln \sum_{i}^{N(\varepsilon)} P_i(\varepsilon)^q}{\ln(1/\varepsilon)}, \tag{5}$$

where $\varepsilon$ refers to the linear size of boxes for spatial measurement, and "lim" denotes the limit in mathematical analysis. From the general definition of fractal dimension, we can derive three special fractal dimension concepts (Grassberger, 1983). If the order of moment $q=0$, we have capacity dimension

$$D_0 = -\frac{S(\varepsilon)}{\ln \varepsilon} = -\frac{\ln N(\varepsilon)}{\ln \varepsilon}, \tag{6}$$

which is based on the macro state entropy. If $q=1$, according to the well-known L'Hôpital's rule, we have information dimension

$$D_1 = -\frac{H(\varepsilon)}{\ln \varepsilon} = \frac{\sum_{i=1}^{N(\varepsilon)} P_i(\varepsilon) \ln P_i(\varepsilon)}{\ln \varepsilon}, \tag{7}$$

which is based on Shannon's information entropy. If $q=2$, we have correlation dimension

$$D_2 = -\frac{M_2(\varepsilon)}{\ln \varepsilon} = \frac{\ln \sum_{i=1}^{N(\varepsilon)} P_i(\varepsilon)^2}{\ln \varepsilon}. \tag{8}$$

The capacity dimension, information dimension, and correlation dimension represent three basic parameters of fractal spectrums. If $D_0=D_1=D_2$, the system bears simple fractal structure, and the fractal object is termed monofractal or unifractal. If $D_0>D_1>D_2$, the system bears complex fractal structure, and the fractal object is termed multifractals. A fractal dimension possesses two spatial meanings, one is spatial entropy, and the other is spatial autocorrelation coefficient (Table 1).

It can be demonstrated that a fractal dimension is a characteristic value of the corresponding spatial entropy relative to the linear scale of measurement. The inverse function of equation (5) is a negative exponential function such as

$$\varepsilon(q) = \exp(-\frac{M_q}{D_q}) = a \exp(-b_q M_q), \tag{9}$$

in which the proportionality coefficient $a=1$, and the decay coefficient $b_q=1/D_q$. This suggests that fractal dimension $D_q$ is the average value of $M_q$ based on exponential distribution. In theory, $b_q=1/D_q$; in empirical studies, however, we will have

$$D_q = \frac{R^2}{b_q}, \tag{10}$$

where $R$ denotes the coefficient of correlation between $M_q$ and $\ln(\varepsilon)$. For an exponential distribution function $y=y_0\exp(-x/x_0)$, where $x$ and $y$ denote variables, and $x_0$ and $y_0$ refer to parameters, the scale parameter represents the characteristic value of $x$. In statistical analysis, a characteristic value can be reflected by an average value or standard deviation. Equation (9) suggests that capacity dimension is the characteristic value of Boltzmann's macro state entropy, information dimension is the characteristic value of Shannon's information entropy, and correlation dimension is the characteristic value of the second order Renyi entropy.

Table 1 A comparison between capacity dimension, information dimension, and correlation dimension

| Parameter | Basic measurement | Variable | Spatial meaning |
|---|---|---|---|
| **Capacity dimension $D_0$** | Macro state entropy | Categorical variable (0 or 1) | Whether or not a place (box) bears fractal elements |
| **Information dimension $D_1$** | Information entropy | Metric variable (0~1) | How many fractal elements appear at/in a place (box) |
| **Correlation dimension $D_2$** | Correlation entropy | Categorical or metric variable (0 or 1, or 0~1) | If a place bears fractal elements, how much is the probability of finding other fractal elements within certain radius from the place (box) |

## 2.2 The equivalence of entropy ratio to fractal dimension ratio

For a fractal system, spatial measurement depends on spatial scale, say, the linear size of boxes. Different scales result in different measurement values. In this case, entropy is not a certain quantity, but fractal dimension as a characteristic value of entropy is theoretically determinate. In order to further connect spatial entropy and fractal dimension, we need the concept of maximum entropy and the maximum dimension. By the functional box-counting method, the maximum

entropy can be define by

$$M_{\max} = \ln N_T, \quad (11)$$

where $N_T$ refers to the total number of boxes, including empty boxes and nonempty boxes, and $M_{\max}$ to the maximum entropy, including the maximum macro state entropy $S_{\max}$ (for $q=0$) and maximum information entropy $H_{\max}$ (for $q=1$). Accordingly, the maximum dimension can be defined as

$$D_{\max} = -\frac{M_{\max}(\varepsilon)}{\ln \varepsilon} = -\frac{\ln N_T(\varepsilon)}{\ln \varepsilon}, \quad (12)$$

where $D_{\max}$ denotes the maximum box dimension. In theory, we have $D_{\max}=d=2$ for urban space.

In geo-spatial analysis, the *entropy ratio* can be employed to replace entropy for the sake of lessening the influence of measurement scale (e.g., linear size of boxes). Entropy ratio can also be termed *entropy quotient*. Based on functional box-counting method and Renyi entropy, the general entropy ratio is

$$J_q(\varepsilon) = \frac{M_q(\varepsilon)}{M_{\max}(\varepsilon)} = -\frac{\ln \sum_{i=1}^{N(\varepsilon)} P_i(\varepsilon)^q}{(q-1)\ln N_T(\varepsilon)}, \quad (13)$$

Accordingly, the *redundancy* can be defined as

$$Z_q(\varepsilon) = 1 - J_q(\varepsilon) = \frac{I_q(\varepsilon)}{M_{\max}(\varepsilon)} = 1 - \frac{M_q(\varepsilon)}{M_{\max}(\varepsilon)}, \quad (14)$$

where $I_q(\varepsilon)=M_{\max}(\varepsilon)-M_q(\varepsilon)$ is termed *information gain*, indicating the difference between the actual entropy and the maximum entropy. The information gain is equivalent to the $H$ quantity in the theory of dissipative structure (Prigogine and Stengers, 1984). If the moment order $q=1$, all these concepts, the entropy ratio, redundancy, and information gain, will return to the common expressions based on Shannon's information entropy (see Batty, 1974; Batty *et al*, 2014).

The three measurements based on entropy can be applied to generalized fractal dimension. For example, from the Boltzmann's equation of fractal dimension growth, we can derive the fractal dimension ratio such as $J_q^*=D_q/D_{\max}$ (Chen, 2012). The ratio of actual fractal dimension to the maximum fractal dimension can also be termed *fractal dimension quotient*. For the generalized dimension, equation (5) divided by equation (12) yields

$$J_q^* = \frac{D_q}{D_{\max}} = -\frac{\ln \sum_{i=1}^{N(\varepsilon)} P_i(\varepsilon)^q}{(q-1)\ln N_T(\varepsilon)}, \tag{15}$$

which suggests that the entropy ratio is equal to the fractal dimension ratio. Comparing equation (13) and equation (15) shows that $J_q(\varepsilon)=J_q^*$. Accordingly, we can define a *dissimilarity index* based on the fractal dimension ratio, and the expression is

$$Z_q^* = 1 - J_q^* = \frac{I_q^*}{D_{\max}} = 1 - \frac{D_q}{D_{\max}}, \tag{16}$$

in which $I_q^*=D_{\max}-D_q$ can be termed *fractal dimension gain*, indicating the difference between the actual fractal dimension and the maximum fractal dimension. Comparing equation (14), equation (15) and equation (16) shows that $Z_q(\varepsilon)=Z_q^*$. This suggests that the dissimilarity index based on fractal dimension is equal to the redundancy index based on entropy. In other words, the ratio of actual entropy to the maximum entropy is equal to the ratio of actual fractal dimension to the maximum fractal dimension. If the moment order $q=0$, all these measurements, the fractal dimension ratio, dissimilarity index, and information dimension gain, will return to the simple expressions based on capacity dimension. The corresponding relationships between spatial entropy and fractal dimension can be tabulated as below (Table 2). In theory, $D_{\max}=d=2$, and thus $I^*$ denotes the scaling exponent of density distribution. For capacity dimension, according to equation (6), the nonempty box number is

$$N(\varepsilon) = \varepsilon^{-D_0}. \tag{17}$$

The corresponding total box number is $N_T(\varepsilon)=1/\varepsilon^2$. Thus, the density distribution function is

$$\rho(\varepsilon) = \frac{N(\varepsilon)}{N_T(\varepsilon)} = \varepsilon^{2-D_0} = \varepsilon^{I_0^*}. \tag{18}$$

Apparently, given $D_{\max}=d=2$, it follows that the scaling exponent $2-D_0=I_0^*$. The measurement can be regarded as a potential index of urban development.

Table 2 A comparison between entropy measurements and fractal dimension measurements

| Type | Spatial entropy | | Fractal dimension | |
| --- | --- | --- | --- | --- |
| **Spatial homogeneity** | Entropy | $M_q$ | Dimension | $D_q$ |

| | Entropy ratio (quotient) | $J_q = \dfrac{M_q}{M_{\max}}$ | Dimension ratio (quotient) | $J_q^* = \dfrac{D_q}{D_{\max}}$ |
|---|---|---|---|---|
| **Spatial heterogeneity** | Information gain | $I_q = M_{\max} - M_q$ | Dimension gain | $I_q^* = D_{\max} - D_q$ |
| | Redundancy | $Z_q = 1 - \dfrac{M_q}{M_{\max}}$ | Dissimilarity index | $Z_q^* = 1 - \dfrac{D_q}{D_{\max}}$ |

**Note**: In this table, $D \ne H$, $I \ne I^*$, but $J = J^*$, $Z = Z^*$.

## 3. Materials and methods

### 3.1 Cases of regular fractals

Two approaches can be utilized to testify the theoretical inference that the ratio of the actual entropy to the maximum entropy is equal to the ratio of the actual fractal dimension to the maximum fractal dimension. One is mathematical experiment based on regular fractals, and the other is empirical analysis based on observational data of fractal cities. For simplicity, let's see nine typical geometric objects, including Euclidean figures and regular growing fractals, which bear an analogy with different models of urban form. Starting from the same initiator (a square), but taking different generators with different numbers of fractal copies, we can construct nine types of regular fractals. Several of them are Euclidean shapes which can be treated as the special cases of growing fractals (Figure 1). The fractal dimension ranges from 0 to 2 (Table 3). When the number ratio of fractal copies is less than 5, the fractal dimension is less than 1.5, and the urban space is less filled (undergrowth). The density is too low. When the dimension is less than 1, the compactness ratio is very low. On the contrary, if the number ratio is greater than 8, the fractal dimension will be greater than 1.8, and the urban space will be excessively filled (overgrowth). The density is too high, and the space is too crowded. The two extreme cases are not good for spatial utilization of cities.

Table 3 Fractal dimension values of nine growing fractals (*m*=1, 2, 3, …)

| Figure 1 | Number ratio $(r_n = N_{m+1}/N_m)$ | Scale ratio $(r_s = s_m/s_{m+1})$ | Fractal dimension of form (*D*) | Number of fractal units | Remark |
|---|---|---|---|---|---|
| a | 1 | 3 | 0.000 | $1^m$ | A point (non-fractal) |
| b | 2 | 3 | 0.631 | $2^m$ | Cantor set |
| c | 3 | 3 | 1.000 | $3^m$ | A straight line |
| d | 4 | 3 | 1.262 | $4^m$ | Fractal dust |
| e | 5 | 3 | 1.465 | $5^m$ | Box fractal |
| f | 6 | 3 | 1.631 | $6^m$ | |
| g | 7 | 3 | 1.771 | $7^m$ | |
| h | 8 | 3 | 1.893 | $8^m$ | Sierpinski carpet |
| i | 9 | 3 | 2.000 | $9^m$ | A square (non-fractal) |

**Note**: Figure a, c, and i are in fact Euclidean shapes, which are treated as three special cases of fractal objects.

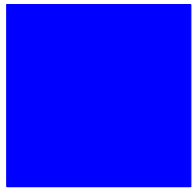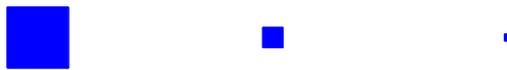

a. $N=1$, $D=\ln(1)/\ln(3)$

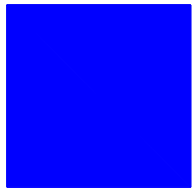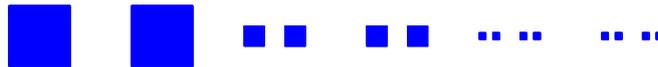

b. $N=2$, $D=\ln(2)/\ln(3)$

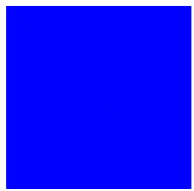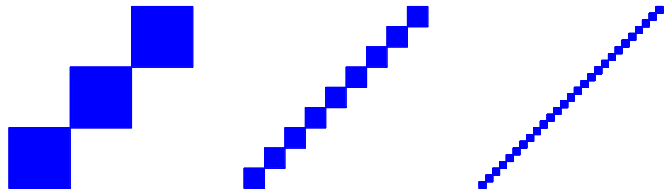

c. $N=3$, $D=\ln(3)/\ln(3)$

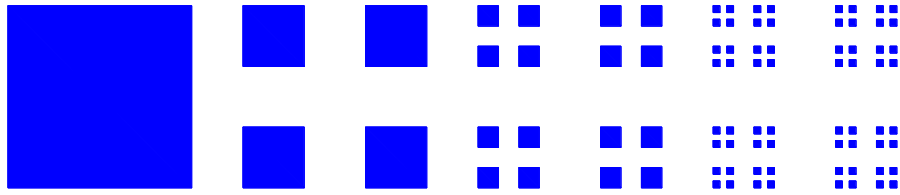

d. *N*=4, *D*=ln(4)/ln(3)

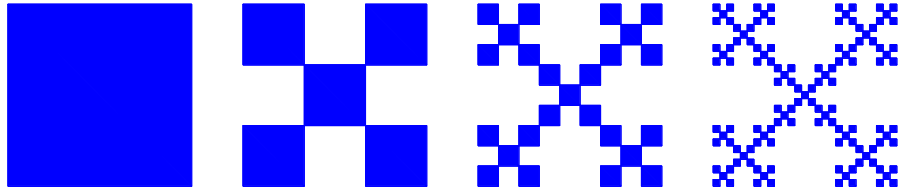

e. *N*=5, *D*=ln(5)/ln(3)

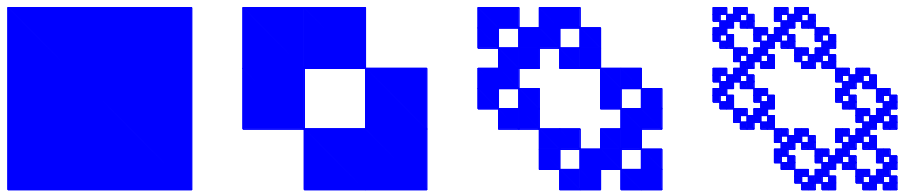

f. *N*=6, *D*=ln(6)/ln(3)

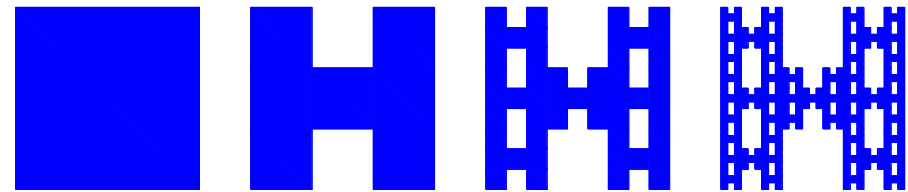

g. *N*=7, *D*=ln(7)/ln(3)

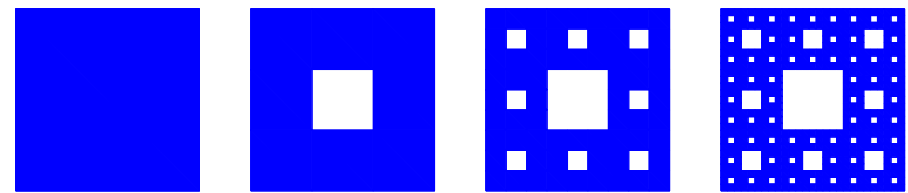

h. *N*=8, *D*=ln(8)/ln(3)

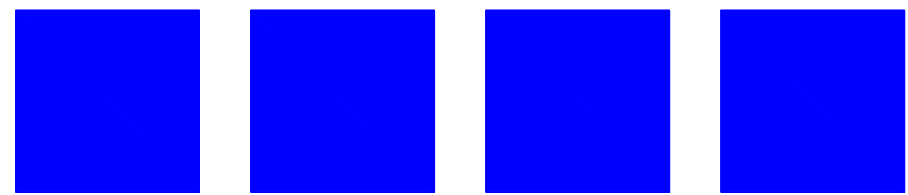

i. *N*=9, *D*=ln(9)/ln(3)

**Figure 1 The generation process of nine regular growing fractals (the first four steps)**

[**Note**: The first one is a point, the third one is a segment of line, and the last one is a square part of plane. The three ones are Euclidean objects, which can be regarded as special cases in a fractal spectrum.]

Fractal dimension is an index of space-filling extent, and this has been illustrated by the regular growing fractals. Accordingly, fractal dimension gain is a potential index of urban land use. Now, let's calculate the spatial information entropy of the growing fractal bodies displayed in Figure 1 by taking Figure 1 (a), (e), and (i) as examples. For simplicity and for comparison with the fractal dimension, we can compute the information entropy based on natural logarithm and nine boxes (Figure 2). The spatial process in Figure 1 (a) represents the absolute concentration and we have only one element (Figure 2 (a)). The information entropy is $H_a=-1*\ln(1)-8*0*\ln(0)$ nat. According to l'Hospital's rule in mathematical analysis, we have $H_a=-\ln(1)=0$ nat. The spatial process in Figure 1 (e) gives the standard growing fractal and we have 5 fractal copies (Figure 2 (b)). The information entropy can be defined as $H_e=-5*(1/5)*\ln(1/5)-4*0*\ln(0)=\ln(5)=1.609$ nat. The spatial process in Figure 1 (i) suggests the complete space filling or absolute even distribution and we have 9 'fractal' units (Figure 2 (c)). The information entropy is $H_e=-9*(1/9)*\ln(1/9)=\ln(9)=2.197$ nat. Generally, given the number of fractal copies $N(\varepsilon)$, it follows that the information entropy

$$H(\varepsilon) = H_m = \ln N(1/3^{m-1}) = (m-1)\ln N(1/3) \quad \text{(nat)}, \tag{19}$$

where $\varepsilon=1/3^{m-1}$. This suggests that the information entropy equals the macro state entropy for the simple fractals. Correspondingly, the fractal dimension is $D=H(\varepsilon)/\ln(\varepsilon)=\ln N(\varepsilon)/\ln(\varepsilon)$, and this indicates that the information dimension equals the capacity dimension for the simple fractal systems. The general expression is

$$D = -\frac{H(\varepsilon)}{\ln \varepsilon} = -\frac{(m-1)\ln N(1/3)}{\ln(1/3^{m-1})} = \frac{\ln N(1/3)}{\ln(3)}. \tag{20}$$

Apparently, the inverse function is $\varepsilon(H)=\exp(-H/D)$, which suggests that fractal dimension $D$ is just the characteristic value (statistical mean) of entropy $H$. This lends support to the formula $D_q=1/b_q$ shown above. These formulae of information entropy and fractal dimension measurement can be applied to other fractal objects shown in Figure 1, and the results are tabulated as blow (Tables 4 and 5).

**Table 4 Comparison of fractal dimension values with information entropy values**

| Figure 1 | Number ratio ($r_n=N_{m+1}/N_m$) | Scale ratio ($r_s=\varepsilon_m/\varepsilon_{m+1}$) | Fractal dimension of form ($D$) | Information entropy for $m=2$ | Dimension entropy ratio |
|---|---|---|---|---|---|

|   |   |   | (*H*) (nat) |   | (*D/H*) |
|---|---|---|---|---|---|
| a | 1 | 3 | 0.000 | 0.000 | -- |
| b | 2 | 3 | 0.631 | 0.693 | 0.91 |
| c | 3 | 3 | 1.000 | 1.099 | 0.91 |
| d | 4 | 3 | 1.262 | 1.386 | 0.91 |
| e | 5 | 3 | 1.465 | 1.609 | 0.91 |
| f | 6 | 3 | 1.631 | 1.792 | 0.91 |
| g | 7 | 3 | 1.771 | 1.946 | 0.91 |
| h | 8 | 3 | 1.893 | 2.079 | 0.91 |
| i | 9 | 3 | 2.000 | 2.197 | 0.91 |

**Table 5 Spatial information entropy and fractal dimension of 9 growing patterns**

| **Type** | **Spatial entropy changing along with scale of measurement** | | | | | | | | **Fractal** |
|---|---|---|---|---|---|---|---|---|---|
| | $1/3^0$ | $1/3^1$ | $1/3^2$ | $1/3^3$ | $1/3^4$ | $1/3^5$ | $1/3^6$ | $1/3^7$ | $1/3^{m-1}$ | **Dimension** |
| a | 0 | 0.000 | 0.000 | 0.000 | 0.000 | 0.000 | 0.000 | 0.000 | $(m-1)\ln(1)$ | 0.000 |
| b | 0 | 0.693 | 1.386 | 2.079 | 2.773 | 3.466 | 4.159 | 4.852 | $(m-1)\ln(2)$ | 0.631 |
| c | 0 | 1.099 | 2.197 | 3.296 | 4.394 | 5.493 | 6.592 | 7.690 | $(m-1)\ln(3)$ | 1.000 |
| d | 0 | 1.386 | 2.773 | 4.159 | 5.545 | 6.931 | 8.318 | 9.704 | $(m-1)\ln(4)$ | 1.262 |
| e | 0 | 1.609 | 3.219 | 4.828 | 6.438 | 8.047 | 9.657 | 11.266 | $(m-1)\ln(5)$ | 1.465 |
| f | 0 | 1.792 | 3.584 | 5.375 | 7.167 | 8.959 | 10.751 | 12.542 | $(m-1)\ln(6)$ | 1.631 |
| g | 0 | 1.946 | 3.892 | 5.838 | 7.784 | 9.730 | 11.675 | 13.621 | $(m-1)\ln(7)$ | 1.771 |
| h | 0 | 2.079 | 4.159 | 6.238 | 8.318 | 10.397 | 12.477 | 14.556 | $(m-1)\ln(8)$ | 1.893 |
| i | 0 | 2.197 | 4.394 | 6.592 | 8.789 | 10.986 | 13.183 | 15.381 | $(m-1)\ln(9)$ | 2.000 |

**Note**: Fractal dimension values represent the characteristic values of the corresponding entropy series, and the linear scales come between 1 (linear size of initiator) and 1/3 (linear size of generator). This implies that the first two steps are very important for fractal dimension measurement.

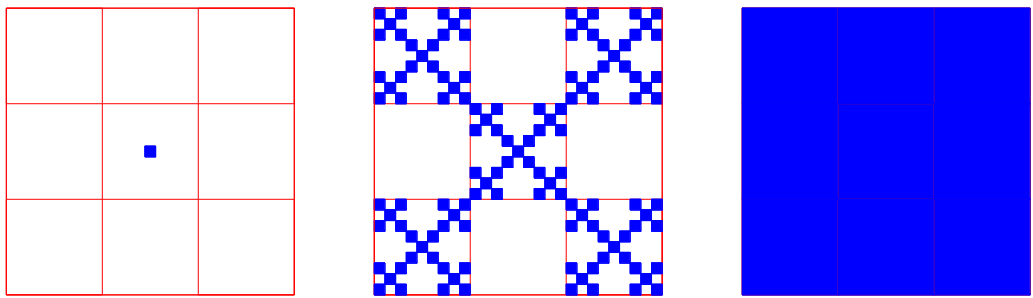

a. Absolute concentration     b. Standard growing fractal     c. Absolute uniformity

**Figure 2 Measuring the spatial entropy of the regular growing fractals with nine boxes**

[**Note**: The first one is a point representing absolute concentration the state of the minimum entropy, and the last one is a square area representing complete space-filling and the state of maximum entropy. The middle one is a growing fractal proposed by Jullien and Botet (1987) and popularized by Vicsek (1989).]

The quotients of entropy and fractal dimension shown above are in fact normalized entropy and fractal dimension. The entropy ratios can be formulated as

$$J_H = \frac{H(\varepsilon)}{H_{max}} = \frac{S(\varepsilon)}{S_{max}} = \frac{\ln N(1/3)}{\ln(9)}, \tag{21}$$

where $H_{max}$ denotes the maximum information entropy. Accordingly, the redundancy is $Z=1-J_H=1-\ln N(1/3)/\ln(9)$. This suggests that, for given number of fractal copies in a generator, $N(1/3)$, the entropy ratios and redundancy are constant. The fractal dimension ratio is

$$J_D = \frac{D}{D_{max}} = \frac{\ln N(1/3)/\ln(3)}{d} = \frac{\ln N(1/3)}{\ln(9)}, \tag{22}$$

in which $d=2$ denotes the maximum dimension value. Accordingly, the dissimilarity index is $Z^*=1-J_D=1-\ln N(1/3)/\ln(9)$. This verifies the inference that a fractal dimension ratio is equal to its corresponding entropy ratio, i.e., $J=J^*$. Based on the values of spatial entropy and fractal dimension, the entropy ratios and fractal dimension ratios of the regular growing fractals can be computed (Table 6). The results show that the normalized spatial entropy exactly equals the normalized dimension of the regular fractals.

**Table 6 Spatial information entropy ratio and fractal dimension ratio of 9 growing patterns**

| Type | Spatial entropy ratios of different fractals | | | | | | | | Fractal dimension rate |
|---|---|---|---|---|---|---|---|---|---|
| | $1/3^1$ | $1/3^2$ | $1/3^3$ | $1/3^4$ | $1/3^5$ | $1/3^6$ | $1/3^7$ | $1/3^{m-1}$ | |
| a | 0.000 | 0.000 | 0.000 | 0.000 | 0.000 | 0.000 | 0.000 | ln(1)/ln(9) | 0.000 |
| b | 0.315 | 0.315 | 0.315 | 0.315 | 0.315 | 0.315 | 0.315 | ln(2)/ln(9) | 0.315 |
| c | 0.500 | 0.500 | 0.500 | 0.500 | 0.500 | 0.500 | 0.500 | ln(3)/ln(9) | 0.500 |
| d | 0.631 | 0.631 | 0.631 | 0.631 | 0.631 | 0.631 | 0.631 | ln(4)/ln(9) | 0.631 |
| e | 0.732 | 0.732 | 0.732 | 0.732 | 0.732 | 0.732 | 0.732 | ln(5)/ln(9) | 0.732 |
| f | 0.815 | 0.815 | 0.815 | 0.815 | 0.815 | 0.815 | 0.815 | ln(6)/ln(9) | 0.815 |
| g | 0.886 | 0.886 | 0.886 | 0.886 | 0.886 | 0.886 | 0.886 | ln(7)/ln(9) | 0.886 |
| h | 0.946 | 0.946 | 0.946 | 0.946 | 0.946 | 0.946 | 0.946 | ln(8)/ln(9) | 0.946 |
| i | 1.000 | 1.000 | 1.000 | 1.000 | 1.000 | 1.000 | 1.000 | ln(9)/ln(9) | 1.000 |

### 3.2 Empirical evidence of Beijing city

Fractal cities can be employed to testify the theoretical results from mathematical derivation and the experiments based on regular fractals. A city fractal bears three basic properties. First, it is a random fractal rather than a regular fractal. A regular fractal object comprises two elements: form and dimension, while a random fractal object comprises three elements: form, chance, and dimension (Mandelbrot, 1977; Mandelbrot, 1982). Second, it is multifractals rather than a monofractal (unifractal). Urban fractal patterns take on clear heterogeneity and multiscale, and cannot be described by a unique fractal parameter (Ariza-Villaverde *et al*, 2013; Chen and Wang, 2013; Murcio *et al*, 2015). Third, it is a prefractal rather than a real fractal. The fractal characters of cities appear within certain range of scales, and fractal dimension can be estimated through scaling range. If the scale is too large or too small, the scaling will break down (Addison, 1997; Bak, 1996; Mandelbrot, 1982). Despite all these problems, fractal geometry can be applied to urban studies, and the effect is encouraging (Batty and Longley, 1994; Chen, 2008; Frankhauser, 1994). Anyway, the city is a scale-free phenomenon, which cannot be effectively described by the traditional mathematical methods based on the ideas of characteristic scales. Fractal theory is a powerful tool of scaling analysis, and can be used to model urban form and growth.

The entropy and fractal indexes can be applied to characterizing the city of Beijing, the capital of China. The form of this city has been demonstrated to bear multifractal properties (Chen and Wang, 2013). Using the functional box-counting method, we can abstract spatial datasets of box sizes and corresponding numbers of nonempty boxes and calculate both Renyi entropy and multifractal parameters. According to the principle of functional box-counting method (Chen, 1995; Feng and Chen, 2010; Lovejoy *et al*, 1987), the single summation formula in equations (1) and (5) should be substituted with double summation formula. The Renyi entropy can be re-expressed as below

$$M_q(\varepsilon) = -\frac{1}{q-1}\log_2 \sum_{i=1}^{n(\varepsilon)}\sum_{j=1}^{n(\varepsilon)} P_{ij}(\varepsilon)^q, \qquad (23)$$

where *i* and *j* denote the number of row and column of box network. For simplicity, we let *i*, *j*=1,2,3,…, $n(\varepsilon)= 1/\varepsilon=2^{m-1}$, in which *m*=1,2,3,…. Accordingly, the generalized dimension can be rewritten as

$$D_q = -\frac{M_q(\varepsilon)}{\log_2 \varepsilon} = -\frac{1}{q-1}\frac{\log_2 \sum_{i=1}^{n(\varepsilon)}\sum_{j=1}^{n(\varepsilon)} P_{ij}(\varepsilon)^q}{\log_2(1/\varepsilon)} = -\frac{\log_2 \sum_{i=1}^{2^{m-1}}\sum_{j=1}^{2^{m-1}} P_{ij}(m)^q}{(q-1)(m-1)}, \qquad (24)$$

in which the natural logarithm is replaced by the logarithm on the base of 2.

It is easy to work out the maximum entropy based on different levels of box networks (grids). Suppose that all the boxes, including the empty boxes and nonempty boxes covering the study area, are evenly filled, thus the entropy reach its maximum value. For given linear size of boxes, $\varepsilon=1/2^{m-1}$, the total box number is

$$N_T(\varepsilon) = n(\varepsilon)^2 = 4^{m-1}. \tag{25}$$

So the maximum entropy is

$$M_{\max}(\varepsilon) = \log_2(4^{m-1}) = 2(m-1). \tag{26}$$

The corresponding unit of information content is bit. In theory, the maximum fractal dimension of fractal cities based on the embedding space with Euclidean dimension $d=2$ is just $D_{\max}=d=2$. Using the maximum entropy $M_{\max}=2(m-1)$ to divide spatial entropy values yields entropy ratios (Table 7). Compared with the entropy values, the entropy ratios are not so dependent on the linear size of boxes. Based on the maximum fractal dimension $D_{\max}=2$, the fractal dimension ratio can be readily calculated. If the linear size of boxes becomes smaller and smaller, then the entropy ratio will become closer and closer to the fractal dimension ratio, that is, $J \to J^*$. Further, we can compute the information gain $I$, fractal dimension gain $I^*$, redundancy $Z$, and dissimilarity index $Z^*$ (Table 8). The values of information gain differ from the values of dimension gain. However, if the linear size of boxes becomes smaller and smaller, then the redundancy will be closer and closer to the dissimilarity index, i.e., $Z \to Z^*$ (Figure 3). By the way, by means of the observational data, it is easy to testify equation (10).

**Table 7 The entropy ratios based on different linear sizes of boxes and the fractal dimension ratios of Beijing's urban form in different years ($m=1,2,3,…,10$)**

| Moment order | Year | $M_q/M_{\max}$ (based on linear size of boxes $\varepsilon=1/2^{m-1}$) | | | | | | | | | $D_q/D_{\max}$ |
|---|---|---|---|---|---|---|---|---|---|---|---|
| | | $1/2^1$ | $1/2^2$ | $1/2^3$ | $1/2^4$ | $1/2^5$ | $1/2^6$ | $1/2^7$ | $1/2^8$ | $1/2^9$ | $1/2^0 \sim 1/2^9$ |
| $q=0$ | 1988 | 1.000 | 1.000 | 1.000 | 0.999 | 0.992 | 0.977 | 0.951 | 0.932 | 0.922 | 0.925 |
| | 1992 | 1.000 | 1.000 | 1.000 | 0.999 | 0.994 | 0.978 | 0.956 | 0.936 | 0.924 | 0.929 |
| | 1999 | 1.000 | 1.000 | 1.000 | 0.999 | 0.996 | 0.987 | 0.970 | 0.955 | 0.945 | 0.950 |
| | 2006 | 1.000 | 1.000 | 1.000 | 0.999 | 0.997 | 0.992 | 0.980 | 0.969 | 0.960 | 0.965 |
| | 2009 | 1.000 | 1.000 | 1.000 | 0.999 | 0.997 | 0.994 | 0.989 | 0.982 | 0.976 | 0.979 |
| $q=1$ | 1988 | 0.984 | 0.894 | 0.900 | 0.908 | 0.913 | 0.913 | 0.908 | 0.906 | 0.906 | 0.905 |
| | 1992 | 0.989 | 0.867 | 0.878 | 0.891 | 0.901 | 0.905 | 0.905 | 0.905 | 0.907 | 0.907 |
| | 1999 | 0.983 | 0.921 | 0.927 | 0.933 | 0.936 | 0.936 | 0.933 | 0.931 | 0.930 | 0.930 |

|  | 2006 | 0.998 | 0.958 | 0.957 | 0.960 | 0.961 | 0.959 | 0.955 | 0.951 | 0.949 | 0.949 |
|  | 2009 | 0.997 | 0.981 | 0.980 | 0.979 | 0.979 | 0.976 | 0.972 | 0.969 | 0.967 | 0.967 |
| *q*=2 | 1988 | 0.968 | 0.819 | 0.836 | 0.859 | 0.876 | 0.887 | 0.892 | 0.896 | 0.901 | 0.902 |
|  | 1992 | 0.978 | 0.784 | 0.814 | 0.845 | 0.866 | 0.880 | 0.889 | 0.896 | 0.901 | 0.904 |
|  | 1999 | 0.964 | 0.867 | 0.883 | 0.898 | 0.910 | 0.916 | 0.920 | 0.922 | 0.925 | 0.926 |
|  | 2006 | 0.996 | 0.924 | 0.930 | 0.938 | 0.944 | 0.945 | 0.945 | 0.945 | 0.946 | 0.945 |
|  | 2009 | 0.994 | 0.965 | 0.966 | 0.968 | 0.969 | 0.968 | 0.965 | 0.964 | 0.963 | 0.963 |

**Note**: When the linear sizes of boxes, $\varepsilon=1/2^{m-1}$, become smaller and smaller, i.e., from 1/2 to $1/2^9$, the entropy ratios become closer and closer to the fractal dimension ratios, i.e., $J=M_q/M_{max} \to J^*=D_q/D_{max}$.

**Table 8 The information/dimension gain and redundancy/dissimilarity index of Beijing's urban form based on functional box-counting method**

| Type | Scale ($\varepsilon$) | $M_{max}$ & $D_{max}$ | Information gain (*I*)/Dimension gain (*I**) | | | | | Redundancy (*Z*)/Dissimilarity index (*Z**) | | | | |
|---|---|---|---|---|---|---|---|---|---|---|---|---|
|  |  |  | 1988 | 1992 | 1999 | 2006 | 2009 | 1988 | 1992 | 1999 | 2006 | 2009 |
| **Entropy parameter** | $1/2^1$ | 2 | 0.033 | 0.022 | 0.035 | 0.004 | 0.006 | 0.016 | 0.011 | 0.017 | 0.002 | 0.003 |
|  | $1/2^2$ | 4 | 0.423 | 0.531 | 0.315 | 0.169 | 0.077 | 0.106 | 0.133 | 0.079 | 0.042 | 0.019 |
|  | $1/2^3$ | 6 | 0.601 | 0.734 | 0.439 | 0.257 | 0.122 | 0.100 | 0.122 | 0.073 | 0.043 | 0.020 |
|  | $1/2^4$ | 8 | 0.735 | 0.870 | 0.539 | 0.322 | 0.166 | 0.092 | 0.109 | 0.067 | 0.040 | 0.021 |
|  | $1/2^5$ | 10 | 0.874 | 0.992 | 0.641 | 0.389 | 0.210 | 0.087 | 0.099 | 0.064 | 0.039 | 0.021 |
|  | $1/2^6$ | 12 | 1.050 | 1.138 | 0.770 | 0.488 | 0.284 | 0.087 | 0.095 | 0.064 | 0.041 | 0.024 |
|  | $1/2^7$ | 14 | 1.285 | 1.328 | 0.941 | 0.632 | 0.389 | 0.092 | 0.095 | 0.067 | 0.045 | 0.028 |
|  | $1/2^8$ | 16 | 1.510 | 1.514 | 1.109 | 0.781 | 0.503 | 0.094 | 0.095 | 0.069 | 0.049 | 0.031 |
|  | $1/2^9$ | 18 | 1.693 | 1.677 | 1.258 | 0.912 | 0.601 | 0.094 | 0.093 | 0.070 | 0.051 | 0.033 |
| **Fractal parameter** | $0$-$1/2^9$ | 2 | 0.190 | 0.187 | 0.140 | 0.101 | 0.067 | 0.095 | 0.093 | 0.070 | 0.051 | 0.033 |

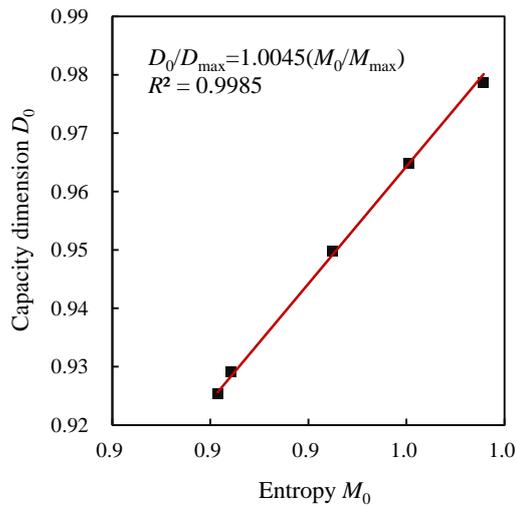
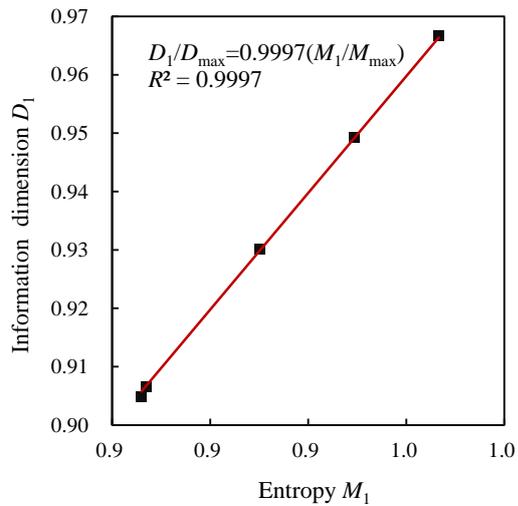

a. Capacity dimension $D_0$ and entropy $M_0$   b. Information dimension $D_1$ and entropy $M_1$

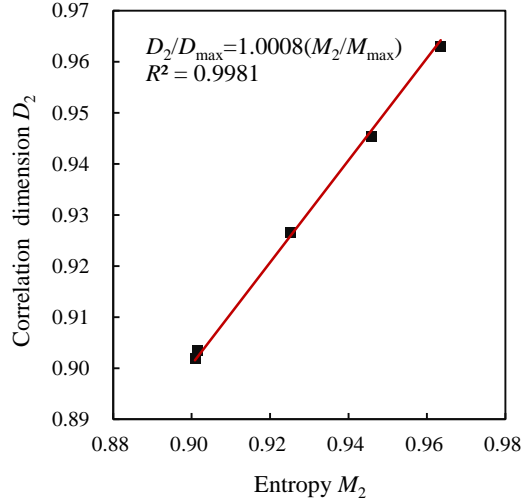

c. Correlation dimension $D_2$ and entropy $M_2$

**Figure 3 The linear relationships between normalized fractal dimension values and the corresponding normalized entropy values (based on the linear size $1/2^9$) of Beijing's urban form (1988-2009)**

## 4. Discussion

The theoretical derivation, mathematical experiments, and empirical analysis suggest that the normalized parameters of spatial entropy are equivalent to the corresponding normalized parameters of fractals. A fractal is in fact a hierarchy with cascade structure, which can be described with a power function or a pair of exponential functions. As a self-similar hierarchy, a fractal object can be modeled by two exponential functions as below:

$$N_m = N_1 r_n^{m-1} = N_0 e^{\ln(r_n)m}, \tag{27}$$

$$S_m = S_1 r_s^{1-m} = S_0 e^{-\ln(r_s)m}, \tag{28}$$

where $m$ denotes the level number of fractal hierarchy or step number of fractal generation ($m=1,2,3,\ldots$), $N_m$ refers to the number of fractal copies of the $m$th step or at the $m$th level, $S_m$ to the linear size of the corresponding fractal copies, $N_1$, $S_1$, $r_n$, and $r_s$ are parameters, among which $N_1=1$ and $S_1=1$ represent the number and size of fractal copy number in initiator, $r_n=N_{m+1}/N_m$ and $r_s=\varepsilon_m/\varepsilon_{m+1}$ are number ratio and size ratio, respectively, $N_0= N_1/r_n$, $S_0= N_1 r_s$. From equations (27) and (28) it follows

$$N_m = \mu S_m^{-D}, \tag{29}$$

where $\mu = N_1 S_1^D$ refers to proportionality coefficient, and $D$ to the fractal dimension. The fractal dimension is the similarity dimension, which can be expressed as

$$D = -\frac{\ln(N_{m+1}/N_m)}{\ln(S_{m+1}/S_m)} = \frac{\ln(r_n)}{\ln(r_s)}, \tag{30}$$

which suggests that $r_n = r_s^D$. The reciprocals of the number ratio logarithm, $1/\ln(r_n)$, and size ratio logarithm, $1/\ln(r_s)$, are two characteristic values. Equation (30) indicates that the fractal dimension is the ratio of two characteristic values. For the simple regular fractals, if fractal copies at the same level do not overlap each other, the similarity dimension equals the box dimension. For Cantor set, $r_n=2$, $r_s=3$, thus the fractal dimension $D=\ln(2)/\ln(3)\approx 0.631$; For Koch curve, $r_n=4$, $r_s=3$, thus the fractal dimension $D=\ln(4)/\ln(3)\approx 1.262$; For Sierpinski gasket, $r_n=3$, $r_s=2$, thus the fractal dimension $D=\ln(3)/\ln(2)\approx 1.585$. The rest can be known by analogy.

The normalized entropy of regular monofractal object does not depend on the linear scale of spatial measurement. Let us examine the entropy value of fractal generator (the second step, i.e., $m=2$). If the Euclidean dimension of the embedding space of a fractal is $d=2$, then $r_n=r_s^d$, and thus the maximum fractal dimension is

$$D_{max} = \frac{\ln(r_s^d)}{\ln(r_s)} = d = 2. \tag{31}$$

That is to say, the fractal dimension comes between topological dimension $d_T$ and the dimension of embedding space $d$ ($d_T \leq D \leq d=2$). If the whole space is filled, the maximum entropy is

$$M_{max} = \ln(r_s^d) = d\ln(r_s) = 2\ln(r_s). \tag{32}$$

The actual entropy of a fractal generator is

$$M_q = S = \ln(r_n). \tag{33}$$

Thus the ratio of actual entropy to the maximum entropy is

$$\frac{M_q}{M_{max}} = \frac{\ln(r_s)}{2\ln(r_n)} = \frac{D}{2} = \frac{D_q}{D_{max}}, \tag{34}$$

which suggests that the entropy ratio equals the fractal dimension ratio. In other words, the ratio of the actual entropy to the maximum entropy is equal to the ratio of the actual fractal dimension to the maximum fractal dimension.

A series of regulars fractals with the same scale ratio have been investigated above. If we check more regular geometric fractals, we will find that different cases lead to the same result. If and only if the scale ratios, i.e., $r_s=\varepsilon_m/\varepsilon_{m+1}$, are identical, the ratio of entropy to fractal dimension ($M_q/D_q$) are constant, otherwise the entropy-dimension ratios are different. However, the ratio of actual entropy to the maximum entropy is equal to the ratio of actual fractal dimension the maximum fractal dimension (Table 9). For the random prefractals in the real world, there are no clearly predetermined scale ratios. The scale ratios can be defined by comparability principle. This implies that we can always find numerical relationships between entropy and fractal dimension in empirical research by means of functional box-counting method.

**Table 9 The entropy values of generators, fractal dimension values, the normalized entropy and fractal dimension of various simple regular fractals**

| Fractal | $r_s$ | $r_n$ | Embedding dimension $d$ | Entropy $M_q$ | Dimension $D_q$ | $M_{max}$ | $D_{max}$ | $J= M_q/M_{max}$ | $J^*= D_q/D_{max}$ |
|---|---|---|---|---|---|---|---|---|---|
| Cantor set | 3 | 2 | 1 | 0.693 | 0.631 | 1.099 | 1 | 0.631 | 0.631 |
| Straght line | 3 | 3 | 1 | 1.099 | 1.000 | 1.099 | 1 | 1.000 | 1.000 |
| Koch curve | 3 | 4 | 2 | 1.386 | 1.262 | 2.197 | 2 | 0.631 | 0.631 |
| Box fractal | 3 | 5 | 2 | 1.609 | 1.465 | 2.197 | 2 | 0.732 | 0.732 |
| Expanded Sierpinski gasket | 3 | 6 | 2 | 1.792 | 1.631 | 2.197 | 2 | 0.815 | 0.815 |
| Sierpinski carpet | 3 | 8 | 2 | 2.079 | 1.893 | 2.197 | 2 | 0.946 | 0.946 |
| Peano curve | 3 | 9 | 2 | 2.197 | 2.000 | 2.197 | 2 | 1.000 | 1.000 |
| Sierpinski gasket | 2 | 3 | 2 | 1.099 | 1.585 | 1.386 | 2 | 0.792 | 0.792 |
| Hilbert space-filling curve | 2 | 4 | 2 | 1.386 | 2.000 | 1.386 | 2 | 1.000 | 1.000 |

The new indexes defined above are mainly normalized measurements. The entropy ratio is actually normalized entropy, the fractal dimension quotient is normalized fractal dimension, the redundancy is normalized information gain, and the dissimilarity index is normalized fractal dimension gain. Based on the Boltzmann's equation of fractal dimension growth of urban form, we obtain a normalized fractal dimension as below (Chen, 2012)

$$D^* = \frac{D_q - D_{min}}{D_{max} - D_{min}}, \qquad (35)$$

where $D_{min}$ refers to the minimum fractal dimension value. Equation (35) is a normalization formula based on the range of fractal dimension. In theory, the Lebesgue measure of a fractal object is zero. This suggests that the minimum dimension can be taken as $D_{\min}=0$. Thus, equation (35) changes to $D^*=D_q/D_{\max}=J^*$. The normalized fractal dimension equals the fractal dimension quotient. Similarly, we can define normalized spatial entropy such as

$$J_q(\varepsilon) = \frac{M_q(\varepsilon) - M_{\min}(\varepsilon)}{M_{\max}(\varepsilon) - M_{\min}(\varepsilon)}, \qquad (36)$$

where $M_{\min}$ denotes the minimum entropy value. Consider two special state of urban form. One is absolutely concentrated to a point, corresponding to the minimum entropy $M_{\min}=0$ (Figure 2a); and the other is absolutely even in a 2-dimensional space, corresponding to the maximum entropy $M_{\max}=2(m-1)$ (Figure 2c). The real form comes between the two extreme cases (Figure 2b). Thus the normalized entropy can be reduced to $J=M_q/M_{\max}$. The theoretical relation between spatial entropy and fractal dimension is as follows

$$\frac{M_q(\varepsilon) - M_{\min}(\varepsilon)}{M_{\max}(\varepsilon) - M_{\min}(\varepsilon)} = \frac{D_q - D_{\min}}{D_{\max} - D_{\min}}. \qquad (37)$$

In practice, equation (37) can be expressed as $J(\varepsilon) \to J^*$, where the arrow "$\to$" denotes "approaches". If the linear size of boxes is smaller enough, say, $\varepsilon < 1/2^{10}$, the arrow will be replaced by an equal sign "=". On the other, 1 minus both sides of equation (37) yields

$$\frac{M_{\max}(\varepsilon) - M_q(\varepsilon)}{M_{\max}(\varepsilon) - M_{\min}(\varepsilon)} = \frac{D_{\max} - D_q}{D_{\max} - D_{\min}}. \qquad (38)$$

If $M_{\min}=0$ and $D_{\min}=0$ as given, then we will have $Z(\varepsilon) \to Z^*$. If the linear size of boxes is smaller enough, we will get $Z(\varepsilon)=Z^*$.

Geographical spatial entropy is always based on Shannon's information entropy. Shannon entropy is conceptually equivalent to thermodynamic entropy (Bekenstein, 2003). There is an analogy between spatial entropy of urban systems and thermodynamic entropy of physical systems. In physics, entropy is a measure of the unavailability of a system's energy to do work. In a closed system, an increase in thermodynamic entropy is certainly accompanied by a decrease in energy availability. For urban studies, spatial entropy is a measure of the unavailability of a city's land for building. In an urbanized area, an increase in information entropy is always accompanied by a decrease in land availability. This implies that spatial entropy can be used to explain and measure spatial utilization of cities. On the other hand, fractal dimension is a space-filling measurement, which reflects the extent of urban land utilization. In order to measure the urban land use, two indexes have been constructed using fractal dimension. One is the *level of space filling* (SFL) of urban form, and the other is the spatial *filled-unfilled ratio* (FUR) of urban growth (Chen, 2012). The former is defined by the normalized fractal dimension

$$D^*(t) = \frac{D(t) - D_{\min}}{D_{\max} - D_{\min}}, \tag{39}$$

where *t* refers to time or year. The latter is defined by the ratio of fractal dimension and fractal dimension gain, that is

$$O(t) = \frac{D^*(t)}{1 - D^*(t)} = \frac{D(t)}{D_{\max} - D(t)}. \tag{40}$$

According to the relationship between spatial entropy and fractal dimension, the SFL is just the entropy ratio, and the FUR is actually the ratio of spatial entropy to information gain. The formulae are as below:

$$J(t) = \frac{M(t)}{M_{\max}} = D^*(t), \tag{41}$$

$$O(t) = \frac{M(t)}{I(t)} = \frac{M(t)}{M_{\max} - M(t)}. \tag{42}$$

These formulae suggest that the essence of space-filling in an urban field is the increase process of spatial entropy. What is more, this implies that we can develop a model to interpret city development by means of entropy and information theory.

The relation between spatial entropy and fractal dimension is revealed by functional box-counting method, and the theory is only suitable for fractal systems. In fact, using the theoretical findings in this article, we can generalize the multifractal parameters and enlarge the sphere of application of multifractal spectrums. Based on equation (34), a pair of global multifractal parameters in a broad sense can be defined as follows

$$D_q = D_{\max} \frac{M_q}{M_{\max}} = \frac{2M_q}{M_{\max}}, \tag{43}$$

$$\tau_q = D_{\max}(q-1)M_q^* = \frac{2(q-1)M_q}{M_{\max}}, \tag{44}$$

where $D_q$ refers to the generalized correlation dimension in the broad sense, and $\tau_q$ to the generalized mass exponent. Then, by means of the Legendre transform, we can derive a pair of local multifractal parameters, including the singularity exponent $\alpha(q)$ and the corresponding local fractal dimension $f(\alpha)$ in a broad sense. The standard multifractal parameters can be only applied to complex systems with multi-scaling processes and patterns. However, the generalized multifractal parameters can be used to model both fractal systems and non-fractal systems, and the functional boxes can be replaced with zonal systems in practical application. After all, a fractal dimension can only be measured for a fractal object, but spatial entropy can be measured for any type of studied objects. The general formulae, equations (43) and (44), are on the base of spatial entropy rather than fractional dimension. The meanings and applications of these generalized parameters will be illustrated in a companion paper (Chen and Feng, 2016).

Compared with the previous studies on spatial entropy based on geographical zoning, this work is based on functional box-counting method. In this way, the measurement of spatial entropy can be standardized and thus associated with box dimension of fractals. The shortcomings of this paper rest with three aspects. First, the local fractal parameters of multifractal systems have not been analyzed by analogy with entropy. Multifractal dimension spectrums comprise both global parameters and local parameters, and this paper is only involved with the global parameters. As indicated above, multifractals bear an analogy with thermodynamics, and, according to Legendre transform, the local fractal dimension is analogous to the entropy (Stanley and Meakin, 1988). Second, spatial entropy has not been connected with radial dimension, which is defined on the base of the area-radius scaling (Frankhauser and Sadler, 1991). Box dimension reflects the patterns of spatial distribution, while radial dimension reflects the relationships between core and periphery of cities (Batty and Longley, 1994; Frankhauser, 1998). For the regular growing fractal, box dimension equals radial dimension (Batty and Longley, 1994). Third, the entropy and dimension of random fractals have not been discussed despite the empirical analysis based on a real fractal city. The problems remain to be solved in future studies.

## 5. Conclusions

Fractal dimension proved to the characteristic of entropy relative to linear scales of spatial measurement. From the theoretical derivation, fractal analyses, and empirical evidence, several clear conclusions can be drawn in this study. The main conclusions are as follows. **First, for the regular simple fractals, the normalized entropy is strictly equal to the normalized fractal dimension. Therefore, the ratio of actual entropy to the maximum entropy is equal to the ratio of actual fractal dimension and the maximum fractal dimension.** If different fractals share the same scale ratio, the ratio of entropy to fractal dimension is a constant, and normalized entropy equals normalized fractal dimension. If different fractals bear different scale ratios, the entropy-dimension ratios will be different, but the normalized entropy value is still equal to the corresponding normalized fractal dimension value. In theory, the initial distribution of a growing fractal system can be defined on the base of a certain state or a point, and the minimum entropy and fractal dimension are zero. **Second, for real complex systems like cities, if the linear scale of spatial measurement such as the linear size of boxes is small enough (e.g., $<1/2^{10}$), the normalized entropy is infinitely approximate to the corresponding normalized fractal dimension.** Complex spatial systems such as cities and network of cities bear the properties of multifractals and random prefractals, which is different from the regular monofractals. However, if we using the functional box-counting method to estimate multifractal parameters, the basic conclusions based on the simple regular fractals are still valid. When the linear size of boxes becomes smaller and smaller, the ratio of actual entropy to the maximum entropy become closer and closer to the ratio of actual fractal dimension to the maximum fractal dimension. **Third, based on the relation between spatial entropy and fractal dimension, the multifractal parameters can be generalized to describe both scaling distributions and the distributions with characteristic scales.** The standard multifractal scaling is only suitable for complex fractal systems. However, if we define a set of parameters based on spatial entropy by analogy with multifractal theory, we will have a set of new formulae similar to the mathematical expressions of multifractal parameters, and these formulae can be employed to describe both fractal cities and non-fractal cities. Thus, the functional boxes will be replaced with systems of zones, and the application area of multifractal theory will be expanded to a degree. It is difficult to clarify all these questions in a few lines of words, and the principle and empirical results will be illuminated and presented in future works.

# Acknowledgements

This research was sponsored by the National Natural Science Foundation of China (Grant No. 41171129). The supports are gratefully acknowledged.